\documentclass[12pt,a4paper]{article}
\usepackage{a4wide}
\usepackage{latexsym}
\usepackage{epsf}
\usepackage{amssymb,amsmath}
\begin{document}
\numberwithin{equation}{section}
\def\be{\begin{equation}}
\def\ee{\end{equation}}
\def\bea{\begin{eqnarray}}
\def\eea{\end{eqnarray}}
\def\pd{\partial}
\def\tr{{\rm tr\,}}
\newcommand{\vev}[1]{\langle\,#1\,\rangle}
\title{\Large\bfseries Minimal strings and Semiclassical Expansion}
\author{\sc C\'esar G\'omez, Sergio Monta\~nez, Pedro Resco\\
\it Instituto de F\'{\i}sica Te\'orica, C-XVI,
Universidad Aut\'onoma de Madrid\\
E-28049-Madrid, Spain}
\date{}
\maketitle
\begin{abstract}
The target space of minimal $(2,2m-1)$ strings is embedded into the phase space of an integrable mechanical model. Quantum effects on
the target space correspond to quantum corrections on the mechanical model. In particular double scaling is equivalent to standard 
uniform approximation at the classical turning points ot the mechanical model. After adding ZZ brane perturbations the quantum target 
remains smooth and topologically trivial. Around the ZZ brane singularities the Baker-Ahkiezer wave function is given in terms 
of the parabollic cylinder function. 
\end{abstract}
\section{Introduction}
In the last years there has been  important progress in the understanding of two
dimensional quantum gravity.
An important ingredient is the definition of D-branes like objects in the context of Liouville theory \cite{FZZ,T,ZZ}, namely
FZZT (or D1) branes and ZZ (or D0) branes.
FZZT branes are one dimensional objects extended in Liouville direction that are
parametrized by the boundary cosmological constant $m_B$, leading
to a natural identification of the moduli space of these branes with the target space of the theory \cite{m}.

For Liouville theory coupled to $(2,2m-1)$ minimal matter and
at lowest order in the string coupling constant the target space defined as the moduli of
FZZT branes, is a Riemann surface \cite{s}.In order  to study all the worldsheet quantum corrections
to this target it  is necessary to use the discrete
(matrix model) description, and to use the correspondence between the FZZT brane
and the double scaling limit of the
macroscopic loop operator in the matrix model \cite{m}.
The ZZ branes in this context correspond to isolated eigenvalues of the matrix model located
out of the Fermi sea.

Our method to study the quantum geometry of this kind of minimal strings
consists in the definition of a map between the brane
amplitudes and an integrable mechanical system\cite{nos}.
By this map the classical target space of the
minimal string (the Riemann surface) goes into a classical curve
in the phase space of the mechanical model.
We conjecture that the quantum gravity effects that change the classical geometry
correspond to the quantum corrections of
the mechanical system. We can proof that
the corrections exactly agree to all orders in the string coupling constant
(Planck constant of the mechanical model) in
the simplest case of the $(2,1)$ minimal string. Moreover by means of this map we
see that the meaning of the double scaling limit of the
matrix model is the quantum resolution of the
singularities that appear at the classical turning points of the mechanical system.

After adding perturbative corrections associated with the
presence of ZZ branes we find that the patern of singularities for the WKB wave function are
unchanged.
This means that the space-time
remains singular even after ZZ perturbative corrections. The quantum uniform approximation
leads, even in the presence of ZZ branes, to a smooth complex plane target space.

\section{Minimal String: Classical and Quantum Target}

\subsection{Classical Target}
Minimal strings are defined as minimal conformal field theories coupled to Liouville
theory
\[
S=S_{m}+S_L
\]
where
\[
S_L=\int \pd\phi^2+QR\phi+me^{2b\phi}
\]
The perturbative expansion of this theory contains only closed surfaces and the string coupling constant is
\[
k\sim e^{Q\phi}
\]

The naive target space for these string theories is the one-dimensional line parametrized by $\phi$
with a region of strong coupling ($Q\phi$ large) where the
perturbative expansion have no sense. In order to probe this
strong coupling region region one can use FZZT branes \cite{FZZ,T}.
Conformal invariance allow us to add a boundary
interaction term
\[
S_\pd =m_B\int_\pd e^{b\phi}
\]
where $m_B$ is the boundary cosmological constant. In the mini-superspace
approximation the FZZT wave function is
\[
\psi(\phi)=\int D\phi D(matter)e^{-S-S_\pd}
\]
where the functional integral is over the disk with the appropiate boundary conditions and the measure is such that $\psi=1$ when
the boundary cosmological constant is zero. At leading order
\[
\psi=e^{-m_Be^{b\phi}}
\]
which means that the brane is extended in the Liouville direction and disolves at
\[
\phi_*\sim-\frac{1}{b}\log m_B
\]
thus we can use the tip of the brane as a probe of the strong coupling region and the moduli space of
the brane as a model for the target space of the theory. The geometry and topology of
this target will be given by the FZZT brane amplitude as a function of the
moduli $m_B$.

Consider the minimal string of type $(2,2m-1)$ and the FZZT disk amplitude $D(x)$
as a function of the moduli $m_B=x$. This amplitude have brach cuts as a function of $x$
(considering $x$ as a complex variable) leading to a target space defined by
the corresponding
Riemann surface. This
Riemann surface emerges because we are probing the strong coupling region and the branch points correspond to deep strong coupling.
The surprising thing here is that the naive one-dimensional target parametrized by $\phi$
has been promoted to a two-dimensional target
(the Riemann surface). However, all physical quantities are holomorphic functions on the Riemann surface so the the target remains
one-dimensional in some sense.

One can see that if one defines $y=\pd_xD$, the Riemann surface of the $(2,2m-1)$ string is
defined by the algebraic equation
\[
F(x,y)=T_2(y)-T_{2m-1}(x)=0
\]
where $T_k$ are the Chebyseb polynomials of first kind. Note that we can write this equation as
\[
2y^2=4^{m-1}(1+x)\prod_{n=1}^{m-1}(x-x_n)^2
\]
In this form is clear that the Riemann surface have $m-1$ singular points at $x=x_n$.
This points correspond to the positions of
the posible ZZ branes of the model.

Note that all these results come from a computation at lowest order in the
string coupling, so it is interesting to consider how
quantum (string coupling) corrections modify this result for target space.
\subsection{Quantum Target: Matrix Models}
In order to compute string corrections to the FZZT brane amplitude we
can use the description of the $(2,2m-1)$ minimal string as the double
scaling limit of a $N\times N$ matrix model with free energy
\[
e^Z=\int dMe^{-\tr V(M)/g}
\]

The standard correspondence relates the disk amplitude of the string
side with the expectation value of the macroscopic loop operator
\[
D\rightarrow \vev{W(x)}=\frac{1}{N}\vev{\tr\log (x-M)}
\]
where $x\sim m_B$ is the coordinate of the matrix eigenvalues.
Using this identification we relate $y(x)$
(that comes fron $F(x,y)=0$) with the resolvent of the matrix model
\[
y(x)\rightarrow R(x)=\pd_x\vev{W(x)}
\]

In order to compute the complete brane amplitude we have to consider an arbitrary
number of boundaries in the string amplitude,
which in the matrix model language implies an exponenciation of the macroscopic loop operator
\[
Z_{Brane}\rightarrow \vev{e^{NW}}=\vev{\det(x-M)}
\]

At  lowest order in the string coupling $k$ we get
\[
Z_{Brane}\sim e^{N\vev{W}}\sim e^{D/k}
\]
which agrees with the WKB expansion in the string side.

If we want a finite double scaling limit for $Z_{Brane}$ we have to identify it with the double scaling limit of
\[
\frac{1}{\sqrt{h_N}}e^{-V(x)/2g} \vev{\det(x-M)}
\]
where $V$ is the matrix model potential, $g$ is the matrix coupling and $h_N$ the normalization constant of the orthogonal
polynomial $P_N(l)=l^N+\dots$ of the matrix model. With these definitions, the complete  brane amplitude corresponds to the
Baker-Akhiezer function of the KP hierarchy associated to the minimal string.

An important property of the Baker-Akhiezer function is that it is an entire (single-valued) function of $x$. This means that if we
consider the exact brane amplitude we will not find a Riemann surface at all as the moduli of the brane. We simply find a complex plane.

Note that the double scaling limit implies a zoom at the edge of the eigenvalue distribution of the matrix model. This fact will be
very important when we try to describe such double scaling as a uniform approximation af a mechanical system at the classical
turning points.

\section{The Mechanical Analog}
To organize the quantum corrections of the classical geometry of the minimal string we use an
analogy with a classical
mechanical model. In the last section we review that the exact brane amplitude in the lowest order WKB expansion can be written as
\[
Z_{Brane}\sim e^{D/k}
\]
where $D$ is the disk amplitude and $k$ is the string coupling constant. This expresion is reminiscent to the wave function at
zero order in WKB of a mechanical model of action $S=-iD$ and $\hbar=k$.
This suggest us to define a correspondence
between mechanical models and minimal strings. Let us  explore this analogy in more detail.
\subsection{One-dimensional Mechanical systems}
The classical motion of an integrable mechanical system in phase space $(p,q)$ is restricted to a curve $\gamma$ defined by $p=p(q,E)$.
The reduced  action associated with this curve is
\[
S(q,E)=\int dq p(q,E)
\]

For the $(2,2m-1)$ string we have as data the  disk amplitude $D(x)$. If we define the map
\[
D(x)=iS(q=x,E=0)
\]
the Riemann surface defined by $y(x)$ map to $p(q=x,E=0)$ (up to analytic continuation to complex time in the mechanical system).

As an example consider the classical target of the minimal $(2,1)$ string . The Riemann surface is defined by
\[
y=\sqrt{\frac{x+1}{2}}
\]
which leads to a classical mechanical system defined by
\[
p(q,E=0)=\sqrt{-\frac{q+1}{2}}
\]

In general, for the $(2,2m-1)$ model we have
\begin{equation}\label{classicalriemann}
y=C\sqrt{(1+x)\prod_{a=1}^{m-1}(x-x_a)^2}
\end{equation}
where $C$ is a constant and $x_a>-1$ for all $a$. The classical curve in phase space is defined by
\be\label{c}
p(q,E=0)=C\sqrt{-(q+1)\prod_{a=1}^{m-1}(q-x_a)^2}
\ee
that implies that we find singularities in the WKB approximation at $x=-1$ (the edge of the double scaled Fermi sea) and at $x=x_a$ (the position of the ZZ branes).
Using this map, the leading WKB aproximation to the wave function of the brane maps to (fixing $\hbar=k$)
\[
\psi=e^{iS(q,E=0)/\hbar}
\]
that is the leading WKB wave function of the mechanical model.
The basic idea of our conjecture is that we can take seriously the map
and use the standard methods of semiclassical quantum mechanics to study the quantum corrections of the wave function of the brane.

Notice that the mechanical systems we are defining corresponds to the double scalling of the matrix model. In fact it is posible
to define a mechanical model analogous to the matrix model associated with the minimal string (see \cite{nos} for details) such that the
classical curves that we are finding here correspond to a zoom near the edge of the eigenvalue distribution of the matrix model potential
that is in correspondence with the classical turning point of the analogous mechanical model.

\subsection{Semiclassical Approximations}

The basic problem of WKB approximation is that it is not well defined near the classical turning points of the motion (branch
points in the Riemann surface). To see this let us  consider the next term in the WKB expansion of the mechanical system
\[
\psi\sim\frac{1}{\sqrt{p}}e^{iS/\hbar}
\]
In the classical turning points $p=0$ the wave function has a divergence. To avoid this problem we use the standard method of
uniform aproximation \cite{b2}. Let us consider the curve associated with the $(2,1)$ model.
The equation satisfied by the wave function is
\[
\psi''-\frac{q+1}{2\hbar^2}\psi=0
\]
The solution of this equation is the Airy function
\[
\psi=Ai(\frac{q+1}{2^{1/3}\hbar^{2/3}})
\]
that agrees with the Baker-Akhiezer function of the string model!.

With this example we see that the uniform aproximation is in some sense the same that double scaling. In double scaling we begin with a
critical point in the matrix potential and perform a zoom around it to find the corresponding expresion for the brane amplitude. In
the mechanical model side we make a zoom around the clasical turning point and find an ``effective'' classical curve around it.
The rules that mach the wave functions in both sides of the turning point give us the correct wave function.

This example present Stokes phenomenon. To study this phenomenon in detail consider the integral representation of the Airy function
\[
Ai(q/\hbar^{2/3})=\int du e^{iuq/\hbar^{2/3}+iu^3/3}
\]
In the semiclassical ($\hbar\rightarrow 0$) limit this integral can be evaluated in the saddle point approximation. It is easy to see that
for $arg(q)>2\pi/3$ two saddles contribute to the integral with imaginary exponential. This gives us an oscilatory behavior in that
region. For $arg(q)<2\pi/3$ we have also two extrema of the exponential, with real exponential contribution, but one of then is a
maximum so it  does not contribute in the saddle point approximation. In this region we finally
find that only one saddle contributes and gives us
the exponential supression after the classical turning point. This phenomenon tell us that the transition from a contribution of two
saddles to a contribution of only one of then is smooth and there is no singularity in the classical turning point. This means also that,
because the Airy function have no branch points in the complex $q$ plane, there is  no Riemannn surface structure in the target
space of the minimal string and the final target space is a complex plane with all physical objects holomorphic over the plane.

\subsection{ZZ-branes and Resolution of Space-Time Singularities}

For models with $m>1$ the situation is similar but corrections asociated with ZZ brane states appear. At first order in the string
coupling constant, the presence of ZZ branes modify the Classical Riemann (\ref{classicalriemann}) surface by a term \cite{s2}
\[
\delta y^2= 2^{2m-3}g_s\sum_a N_a\sqrt{1+x_a}\prod_{l\neq a}(x-x_l)
\]
where the $x_a$  corresponds to the different
positions of the ZZ branes (singularities in the classical Riemann surface). $N_n$ is the number of ZZ branes located at each
singular point. The singularities of the associated WKB wave function before adding
the perturbative ZZ corrections are in $x=-1$
(that do not correspond to a singularity of the classical Riemann surface)
and at $x=x_a$ (associated with the space-time singularities).
One expect that the extra term split the singularities of the classical Riemann
surface but in the semiclassical approximation one
find that after the ZZ brane correction  the WKB wave funcion of the associated
mechanical model is modified only by a multiplicative factor
\[
\prod_a(\frac{\sqrt{1+x}+\sqrt{x-x_a}}{\sqrt{1+x_a}})^{2^{m-3/2}N_a}
\]
This factor does not change the singular behavior of the wave function and take a constant (and diferent to zero) value at the
singularities .
In the string theory language this means that the singularities of the
classical Riemann surface remains present in the semiclassical (perturbative) expansion. Only the full non-perturvative effects can smooth
the singularities. To study the uniform aproximation we have to focus at the singular poins (that correspond to classical turning
points in the model). For the $(2,2m-1)$ we have $m$ singular points. One of then is located al $x=-1$ and near this point
the momentum behaves as
\[
p^2\sim 1+x
\]
as in the $(2,1)$ case. This implies that near this turning point the wave function behaves as an Airy function. For the other
singular points $x_n$ for $n=1\dots m-1$ the associated classical  momentum behaves as
\[
p^2\sim (x-x_n)^2
\]
This implies that the Airy function is not valid to implement the uniform aproximation near these points. The WKB wave function near
this points have the form
\[
\psi\sim \frac{e^{\pm a(x-x_n)^2/2\hbar}}{(x-x_n)^{1/2}}
\]
Using the standard rules of quantum mechanics and the maching conditions\cite{b2} one can see that the correct form of the
wave function near this turning point is given by\footnote{ $D_{-1/2}$ represents the parabolic cylinder function that we
define in the next section.}
\[
D_{-1/2}(-\frac{x-x_n}{(2\hbar)^{1/2}a^{-1/4}})
\]
and that there is a change in the exponential behavior at the turning point that implies that a negative exponential  before the
turning point matches with a positive exponential behavior after the turning point and viceversa. This behavior agree with the fact
that the models with $m$
even are unstable in a nonperturbative sense, because the asociated wave function presents a singularity at $x\sim \infty$. Note that
this uniform aproximation is the same that we have to use if we forget the ZZ corrections. This is because near this points the extra
prefactor is a constant phase in the wave function.

To see how these ideas work in full detail let us consider the example of only one type of ZZ branes: the $(2,3)$ model.
\subsection{Semiclassical Aproximation in the $(2,3)$ model}
For this model we have that the classical Riemann surface is given by
\[
2y^2=1+T_3(x)=4(1+x)(x-x_1)^2
\]
where $x_1=1/2$. This form implies that we have only one singular point in the Riemann surface at $x=x_1$. The relation between singular
points and ZZ states \cite{s} implies that there is only one type of ZZ branes that are located at $x_1$. To study the effect of
the presence of ZZ branes on the Riemann surface note that the annulus amplitude between the FZZT and $N$ ZZ branes at $x_1$
modify the Riemann surface\cite{s2} by
\[
\delta y^2=2g_s\sqrt{1+x_1}N
\]
where $g_s$ is the string coupling constant. Using the mechanical analog we find that the wave function satisfies the
quantum corrected equation (note that $\hbar=g_s$ in the mechanical analogy)
\[
\hbar\psi''=-(p^2+\hbar\delta p^2)\psi
\]
where
\[
p^2=-2(1+x)(x-x_1)^2
\]
and
\[
\delta p^2=-2N\sqrt{1+x_1}
\]
If we solve the quantum corrected wave equation in the WKB approximation we find that
\[
\psi\sim\frac{e^{\pm i\int\chi/\hbar}}{\sqrt{\chi}}(\frac{\sqrt{1+x}+\sqrt{x-x_1}}{\sqrt{1+x_1}})^{\pm N\sqrt{2}}
\]
where
\[
\chi=\sqrt{2}\sqrt{-1-x}|x-x_1|
\]
Near $x=-1$ this function is singular but one can use the Airy function to remove the singularity and uniformize it. The behavior of
the wave function for $x\sim-1$ and $x<-1$ is
\[
\psi\sim \frac{e^{\pm i \sqrt{8}|1+x_1|(-1-x)^{3/2}/3\hbar}}{\sqrt{\sqrt{2}|1+x_1|\sqrt{-1-x}}}i^{\pm N\sqrt{2}}
\]
and for $x>-1$
\[
\psi\sim \frac{e^{\pm  \sqrt{8}|1+x_1|(1+x)^{3/2}/3\hbar}}{\sqrt{\sqrt{2}|1+x_1|\sqrt{1+x}}}i^{\pm N\sqrt{2}}e^{-i\pi/4}
\]
These kinds of WKB approximations at $x\sim -1$ implies that the correct behavior at $x=-1$ is given by the Airy function
\[
\psi\sim Ai((x+1)\frac{2^{1/3}(1+x_1)^{2/3}}{\hbar^{2/3}})
\]
that give us the correct  exponential supression for $x>-1$.

Near $x=x_1$ we find that the correction introduced by the ZZ branes do not remove the singularity at $x=x_1$ and the behavior of
the WKB wave function is given by
\[
\psi\sim \frac{e^{- \sqrt{2}|1+x_1|(x-x_1)^{2}/2\hbar}}{\sqrt{\sqrt{2}|x-x_1|\sqrt{1+x_1}}}e^{i\pi/4}
\]
for $x<x_1$ (note that the previous regularization using Airy fixes the negative exponential behavior), and
\[
\psi\sim \frac{e^{\pm \sqrt{2}|1+x_1|(x-x_1)^{2}/2\hbar}}{\sqrt{\sqrt{2}|x-x_1|\sqrt{1+x_1}}}e^{i\pi/4}
\]
for $x>x_1$. The WKB approximation is singular but if we use the uniform approximation \cite{b2}for this case one finds that
the correct behavior of the wave function near $x=x_1$ is
\[
\psi\sim 2^{1/4}\frac{\hbar^{1/4}}{(2+2x_1)^{1/8}}D_{-1/2}(-\frac{(2+2x_1)^{1/4}}{\sqrt{2\hbar}}(x-x_1))e^{i\pi/4}
\]
where $D_{-1/2}(-x/\sqrt{2})$ is the parabolic cylinder function solution of the equation
\[
\frac{d^2}{dx^2}\psi-x^2\psi=0
\]
This uniformization fixes the behavior for $x>x_1$ to
\[
\psi\sim \frac{e^{+ \sqrt{2}|1+x_1|(x-x_1)^{2}/2\hbar}}{\sqrt{\sqrt{2}|x-x_1|\sqrt{1+x_1}}}e^{i\pi/4}
\]
that reflects the nonperturbative unstability of the model.
This construction prove that the uniform approximation near the turning points kills the branch points that appear in the different
asymptotic expansions. This fact is clear in the quantum mechanical sense because, for a single-valued potential, the solution of the
one-dimensional Schoedinger equation have to be single valued.
\section{Conclusions}
We  conjecture that the complete amplitude of the FZZT brane of the $(2,2m-1)$ minimal models (that corresponds with the Baker-Akhizer
function of the KP hierarchy associated with the closed minimal string) can be related to the wave function of certain
mechanical model. Using the matrix model realization of the minimal string we find a relation between double scaling limit
and the uniform approximation of wave functions in a mechanical model. We also proof exact agreement for the case of topological
gravity. The  perturbative ZZ brane corrections do not open a new brach cut at the singularities of the classical Riemann surface and one
need the full nonperturvative uniformization to remove the classical space-time singularities.


\end{document}